\documentclass[aps,prl,reprint,superscriptaddress,nofootinbib]{revtex4-2}
\usepackage{local}
\newcommand{\brst}{\abbrev{BRST}}
\newcommand{\RF}{\abbrev{RF}}
\newcommand{\intx}{\int_x}

\begin{document}

\preprint{TTK-26-09}
\title{The perturbative Ricci flow in gravity}

\author{Robert V. Harlander}
\email{robert.harlander@rwth-aachen.de}
\affiliation{TTK, RWTH Aachen University, 52067 Aachen, Germany}

\author{Yannick Kluth}
\email{yannick.kluth@utoronto.ca}
\affiliation{Department of Physics, University of Toronto, ON M5S 1A7, Canada}

\author{Jonas T. Kohnen}
\email{jonas.kohnen@rwth-aachen.de}
\affiliation{TTK, RWTH Aachen University, 52067 Aachen, Germany}

\author{Henry Werthenbach}
\email{henry.werthenbach@rwth-aachen.de}
\affiliation{TTK, RWTH Aachen University, 52067 Aachen, Germany}

\begin{abstract}
We develop a perturbative formulation of the Ricci flow in gravity. Following steps analogous to the gradient flow in \qcd, we supplement the usual Feynman rules for perturbative gravity by flowed propagators and vertices as well as graviton flow lines which describe the evolution of gravity along the Ricci flow. By calculating vacuum expectation values of a number of independent operators at the two-loop level, we derive the required counterterms of the flowed action. Our results allow us to define a Ricci-flow based renormalization scheme for Newton's constant $\gnewton$. Studying its renormalization group behavior, we recover a non-Gaußian fixed point in accordance with well-known non-perturbative considerations.
\end{abstract}

\maketitle


\prlSection{Introduction} Gravity still defies a unified treatment with the
other three known fundamental interactions, which are described by the Standard
Model to a high level of precision~\cite{ParticleDataGroup:2024cfk}. The main
reason is that gravity, due to Newton's constant having negative mass
dimension, is not Dyson renormalizable in four dimensions \cite{Dyson:1952tj}.
Subtracting all \uv\ divergences requires an infinite tower of counterterms
beyond the Einstein-Hilbert action
\cite{Goroff:1985sz,Goroff:1985th,vandeVen:1991gw}. However, there are a number
of indications that gravity could have a non-trivial \uv\ fixed point which
would make it asymptotically
safe~\cite{Weinberg:1976xy,Parisi:1977uz,Weinberg:1980gg,Reuter:1996cp,Ambjorn:2012jv}. 

Most of the relevant theoretical searches for such a fixed point are based on
the non-perturbative \frg\ \cite{Wetterich:1992yh}. Even though those studies
give rise to a fixed point with near-perturbative properties
\cite{Falls:2014tra,Falls:2017lst,Eichhorn:2018akn,Eichhorn:2018ydy,Kluth:2020bdv,Baldazzi:2023pep}, perturbative
approaches face challenges due to the impact of power divergences which must be
captured to find evidence for a fixed point~\cite{Kluth:2024lar}.
Mass-independent schemes like (modified) minimal subtraction (\msbar) are
oblivious to such contributions, which means that non-trivial fixed points can
be hidden~\cite{cckl}. While non-minimal renormalization schemes can be
defined that do support a non-Gaußian fixed point in gravity due to their
sensitivity to power divergences even in perturbation
theory~\cite{Codello:2006in,Niedermaier:2009zz,Niedermaier:2010zz,Martini:2021slj,Kluth:2024lar,Falls:2024noj},
those often face additional challenges. On the one hand, loop approximations
derived from the \frg\ involve complex truncations to capture relevant effects
beyond one-loop and are further complicated from breaking quantum diffeomorphism
invariance by the regulator~\cite{Ebert:2007gf,Pawlowski:2023gym}. Other
perturbative approaches frequently rely on renormalizing two-dimensional gravity
\cite{Kawai:1989yh,Jack:1990ey,Kawai:1992np,Kawai:1993fq,Kawai:1993mb,Kluth:2024lar,Falls:2024noj},
which requires careful distinction of \uv\ poles from kinematic poles that arise
from the propagator due to the topological nature of gravity in $d = 2$
\cite{Martini:2022sll}.

In this paper, we pursue a novel approach that maintains \brst\ invariance and
is sensitive to power divergences without relying on gravity in $d = 2$. This is
achieved by the Ricci flow, which describes the evolution of the graviton field
along an artificial trajectory towards a metric of constant curvature.
Originally introduced by Hamilton~\cite{Hamilton:1982} as a topological concept,
the Ricci flow was famously used by Perelman in his proof of the Poincar\'e
conjecture~\cite{Perelman:2003uq,Perelman:2006un,Perelman:2006up}. Our
motivation, however, comes from quantum gauge theories, where a similar concept,
called the \textit{gradient flow}~\cite{Luscher:2010iy,
Luscher:2011bx,Luscher:2013cpa}, has been used very successfully in the context
of lattice gauge theory, be it for smearing operations, scale setting, the
renormalization of composite operators, etc. The gradient flow can be viewed as
a renormalization scheme which is accessible both from the lattice and within
perturbation theory. In fact, it was realized early-on that it allows for a
definition of the renormalized strong coupling which could form a practical
bridge between the hadronic scale $\Lambda_\text{\qcd}$ and the perturbative
theory parameter $\alpha_s(M_Z)$~\cite{Luscher:2010iy}. It has also proven
successful for the non-perturbative calculation of matrix elements relevant for
the determination of parton densities~\cite{Harlander:2025qsa,Francis:2025rya,Francis:2025pgf},
flavor observables~\cite{Black:2026rbz,Black:2026dzp}, or the energy-momentum
tensor~\cite{Iritani:2018idk}. 

With this in mind, we develop the perturbative description of the Ricci flow in
this paper. The reader familiar with the perturbative gradient flow will
recognize the close parallels to our procedure. After defining the
flowed metric and the flowed quantities derived from it, we supplement the
Feynman rules of the Einstein-Hilbert action by those for the flowed fields.
Subsequently, we evaluate the counterterms which are needed to render the
corresponding Green's functions finite through two-loop level. We then suggest a
renormalization scheme for Newton's coupling based on the Ricci flow. Its
$\beta$ function indeed develops a zero at a non-vanishing, but perturbative
value of the coupling.

\prlSection{Einstein-Hilbert gravity}Our starting point is the Einstein-Hilbert
action for a metric $g_{\mu \nu}$ in $d = 4 - 2 \varepsilon$ space-time
dimensions and with Euclidean signature,\footnote{Similarly to the perturbative
gradient flow for \qcd, we assume a Euclidean signature to avoid obstacles that
arise for flow equations in Lorentzian signatures.}
\begin{equation}
  S_{\text{EH}} = - \frac{2}{\kappa^2} \intx \sqrt{g} R \, ,
    \label{eqn:SEH}
\end{equation}
with $\kappa^2=32\pi\gnewton$ and $\gnewton$ Newton's coupling, $R$ the Ricci scalar, and $\intx=\int\dd^dx$
the integral over $d$ dimensional space-time. Here and in the following, we will
assume a vanishing cosmological constant. We introduce the graviton field
$h_{\mu\nu}$ as the fluctuation of the full metric around a constant flat
background,
\begin{equation}
    g_{\mu \nu}(x) = \delta_{\mu \nu} + h_{\mu \nu}(x) \,.
    \label{eqn:metric_expansion}
\end{equation}
For the perturbative quantization of \cref{eqn:SEH}, we add a gauge-fixing and a
corresponding ghost term to the action,
\begin{equation}\label{eq:56:90}
  \begin{aligned}
    S = S_\text{EH} + S_\text{gf} + S_\text{gh}\,,
  \end{aligned}
\end{equation}
with
\begin{equation}
    S_{\text{gf}} = \intx \, \frac{1}{\alpha \kappa^2} \delta^{\mu \nu} F_\mu F_\nu \, ,\quad
    S_{\text{gh}} =  \intx\, \bC^\mu \frac{\delta F_\mu}{\delta h_{\rho \sigma}} \nabla_\rho C_\sigma \, ,
    \label{eqn:Sgh}
\end{equation}
with $\alpha$ a gauge-fixing parameter, and $C^\mu$, $\bC^\mu$ the Faddeev-Popov ghost and
anti-ghost fields, respectively. A simple linear gauge-fixing is chosen,
\begin{equation}
    F_\mu = \partial^\nu h_{\mu \nu} + \beta \delta^{\rho \sigma} \partial_\mu h_{\rho \sigma} \, ,
    \label{eqn:gauge_functional}
\end{equation}
where $\beta$ is another free gauge parameter. 

In order to obtain \uv\ finite results for physical quantities, \cref{eqn:SEH}
has to be supplemented by counterterms. They can be obtained by renormalizing
the metric as\footnote{The Gauß-Bonnet
term does not play a role in the following due to its topological nature.}
\begin{equation}\label{eq:155:99}
  \begin{aligned}
    g_{\mu\nu} &\to g_{\mu\nu} + \delta g_{\mu \nu}\,,\\ 
\delta g_{\mu\nu} &=  \frac{\gnewton}{4\pi} \,
    \left[c_1 R g_{\mu \nu} + c_2 R_{\mu \nu} \right] \, ,
  \end{aligned}
\end{equation}
where $R_{\mu\nu}$ is the Ricci tensor.  The explicit form of the divergent and
gauge-dependent coefficients $c_1$ and $c_2$ is not needed in our paper; for
$\alpha=-2\beta=1$, it can
be found in Ref.\,\cite{Mandric:2023dmx}.

\prlSection{Perturbative Ricci flow}For any smooth metric $g_{\mu \nu}(x)$, we define a \textit{flowed metric}
$\fg_{\mu \nu} (t,x)$ with
\begin{equation}\label{eq:84:44}
  \begin{aligned}
    \fg_{\mu\nu}(0,x)=g_{\mu\nu}(x) + \delta \hat{g}_{\mu \nu} (x)\,,
  \end{aligned}
\end{equation} 
where $t$ is an auxiliary parameter named the \textit{flow time}, and $\delta
\hat{g}_{\mu \nu} (x)$ collects counterterms that will be discussed in
more detail below. At $t>0$, the flowed metric is defined via the (gauged)
\textit{Ricci-flow equation}
\begin{equation}
    \partial_t \fg_{\mu \nu} = - 2 \fR_{\mu \nu}  + 2\alpha_0 \fnabla_{(\mu} \fF_{\nu)} \equiv \fE_{\mu \nu} \,,
    \label{eqn:gaugefixedfloweq}
\end{equation}

Here, the brackets denote normalized symmetrization, i.e.\
$A_{(\mu\nu)}=(A_{\mu\nu}+A_{\nu\mu})/2$, and $\fR_{\mu\nu}$ is the flowed Ricci
tensor, defined in analogy to the usual (unflowed) Ricci tensor, but with all
metric tensors replaced by flowed ones.  Similar to \cref{eqn:metric_expansion},
we introduce the flowed graviton field $\fh_{\mu \nu}$ as the fluctuation of the
flowed metric $\fg_{\mu \nu}$ around a constant background, $\fg_{\mu \nu}(t,x)
= \delta_{\mu \nu} + \fh_{\mu \nu}(t,x)$.  One can show that the solutions of
\cref{eqn:gaugefixedfloweq} for arbitrary $\alpha_0$ and $\fF$ are related by
$t$ dependent gauge transformations. Therefore, any diffeomorphism invariant
object is independent of $\alpha_0$ and $\fF$.  In analogy to \cref{eqn:gauge_functional},
we choose
\begin{equation}
    \fF_\mu = \partial^\nu \fh_{\mu \nu} + \beta_0 \delta^{\rho \sigma} \partial_\mu \fh_{\rho \sigma} \,,
    \label{eqn:flow_gauge_functional}
\end{equation}
where $\beta_0$ is another arbitrary parameter. 

\cref{eqn:gaugefixedfloweq} can be solved iteratively by  
expanding the right-hand side in powers of the graviton field $\fh_{\mu
\nu}$, which results in a perturbative series for $\fh_{\mu \nu}$ in terms of
convolutions of $h_{\mu\nu}$. Alternatively, one can formulate the Ricci flow as
$(d+1)$-dimensional field theory, with the flow time $t$ acting as an additional
artificial dimension. For this purpose, we define the flowed action
\begin{equation}
    S_{\text{flow}} = S + \int_0^\infty \dd{t} \intx\, \fL^{\mu \nu} \left[ \partial_t \fh_{\mu \nu} - \fE_{\mu \nu} \right] \, .
    \label{eqn:Sflow_h}
\end{equation}
where $\fL^{\mu \nu}(t,x)$ is a Lagrange multiplier field. It is a tensor
density of weight one, which means that the corresponding space-time integral in
\cref{eqn:Sflow_h} does not involve the volume element $\sqrt{\fg}$. 
It can be shown that, by introducing flowed ghosts into \cref{eqn:Sflow_h}, it
can be made invariant under a generalized \brst{} transformation.  Analogous to
the Yang-Mills gradient flow, the flowed ghosts do
not contribute to Green's functions of $\fh_{\mu\nu}$ (or $h_{\mu\nu}$) though, and are
therefore irrelevant for our purposes. 

\prlSection{Feynman Rules} 
The flowed action in \cref{eqn:Sflow_h} supplements the Feynman rules of
regular Einstein-Hilbert gravity by flowed Feynman rules which describe the
evolution towards finite flow time. In particular, with the choice $\alpha_0 = -2\beta_0 = 1$, \cref{eqn:Sflow_h} implies\footnote{The Heaviside function is defined to be $\theta(t) = 0$ at
$t=0$.}
\begin{equation}
  \begin{aligned}
        \langle
            \fh_{\mu \nu} (t, p) \fL_{\rho \sigma} (s, -p) \rangle 
    &\overset{\text{\abbrev{LO}}}{=} \theta(t - s) e^{- (t - s) p^2} \delta_{\mu (\rho} \delta_{\sigma) \nu}\\
    \equiv \raisebox{-1.2em}{\includegraphics[
      width=.15\textwidth]{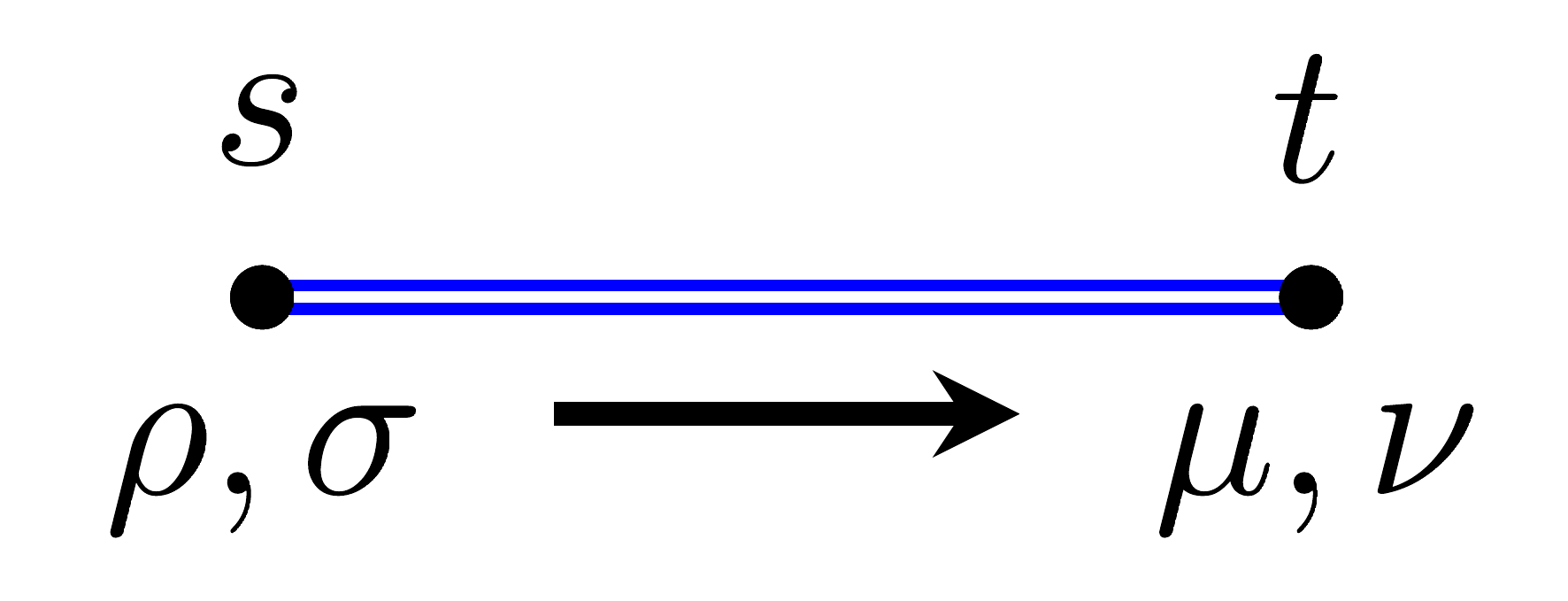}}
    \label{eqn:flowline_kernel}
  \end{aligned}
\end{equation}
This object will be referred to as \textit{(graviton) flow line} in the
following. Since the flow line connects two different fields, it is directional;
this is indicated by the arrow flowing from $\fL$ to $\fh$, or equivalently,
towards increasing flow time. Similarly, \cref{eqn:Sflow_h} implies
\textit{flowed vertices}, all of which are linear in $\fL^{\mu \nu}$, meaning
that they contain exactly one outgoing flow line. The explicit expressions for
these vertices can be obtained in a straightforward manner. Each of them depends
on an independent flow-time parameter which is integrated over from 0 to
$\infty$; however, the Heaviside functions of the flow lines always lead to the
restriction of these flow-time integrals to a finite interval.

Finally, the initial condition of \cref{eq:84:44} is accounted for by
introducing a \textit{flowed graviton propagator},
\begin{equation}
    \begin{split}
        \langle \fh_{\mu \nu} (t, p) \fh_{\rho \sigma} (s, -p) \rangle 
        &\overset{\text{\abbrev{LO}}}{=}\, e^{-(t + s) p^2} 
        \mathcal{P}_{\mu \nu \rho \sigma} (p) \, ,
    \end{split}
    \label{eqn:h_prop_dedonder}
\end{equation}
where $p$ is the momentum, $s$ and $t$ are the flow-time parameters of the
vertices that the propagator is connecting, and $\mathcal{P}_{\mu \nu \rho \sigma} (p)$ is the standard graviton propagator derived from the gauge-fixed
Einstein-Hilbert action of \cref{eq:56:90}, which we provide here for
$\alpha=-2\beta=1$ for the sake of clarity:
\begin{equation}\label{eq:174:48}
  \begin{aligned}
        \mathcal{P}_{\mu \nu \rho
        \sigma} (p) &= 
          \frac{\kappa^2}{p^2} \left( \delta_{\mu (\rho} \delta_{\sigma) \nu}
        - \frac{1}{d - 2} \delta_{\mu \nu} \delta_{\rho \sigma} \right)\,.
  \end{aligned}
\end{equation} 
Similar to the vertices of the Einstein-Hilbert action, flowed vertices may
contain arbitrarily high orders in $\fh_{\mu \nu}$, which are connected to
the rest of the diagram either by incoming flow lines or by flowed propagators.
Nevertheless, at any finite order in the $\gnewton$ expansion, only vertices
with a finite number of fields are required. 

Because the flow time always increases along flow lines, it follows that any
diagram containing closed flow-line loops vanishes identically. This is also the
reason why flowed ghost fields do not contribute to physical observables.
Nevertheless, standard Faddeev-Popov ghosts at vanishing flow-time do still
arise and can give non-vanishing contributions.

Since the flow constitutes a cutoff at finite flow time, \uv\ divergences can
only be associated with fields at vanishing flow time. Therefore, any
divergences beyond those of the regular theory must be related to the initial
condition of \cref{eq:84:44}. For symmetry reasons, the corresponding
counterterms must be of the same form as those in \cref{eq:155:99}, i.e.
\begin{equation}\label{eq:292:48}
  \begin{aligned}
    \delta\hat{g}_{\mu\nu} = 
    \delta g_{\mu\nu}\Big|_{c_i\to \hat{c}_i}\,,
  \end{aligned}
\end{equation}
with coefficients $\hat{c}_i$ to be determined below.

\begin{figure}
  \begin{center}
    \begin{tabular}{cccc}
      \raisebox{1.6em}{%
              \includegraphics[width=.1\textwidth]%
                              {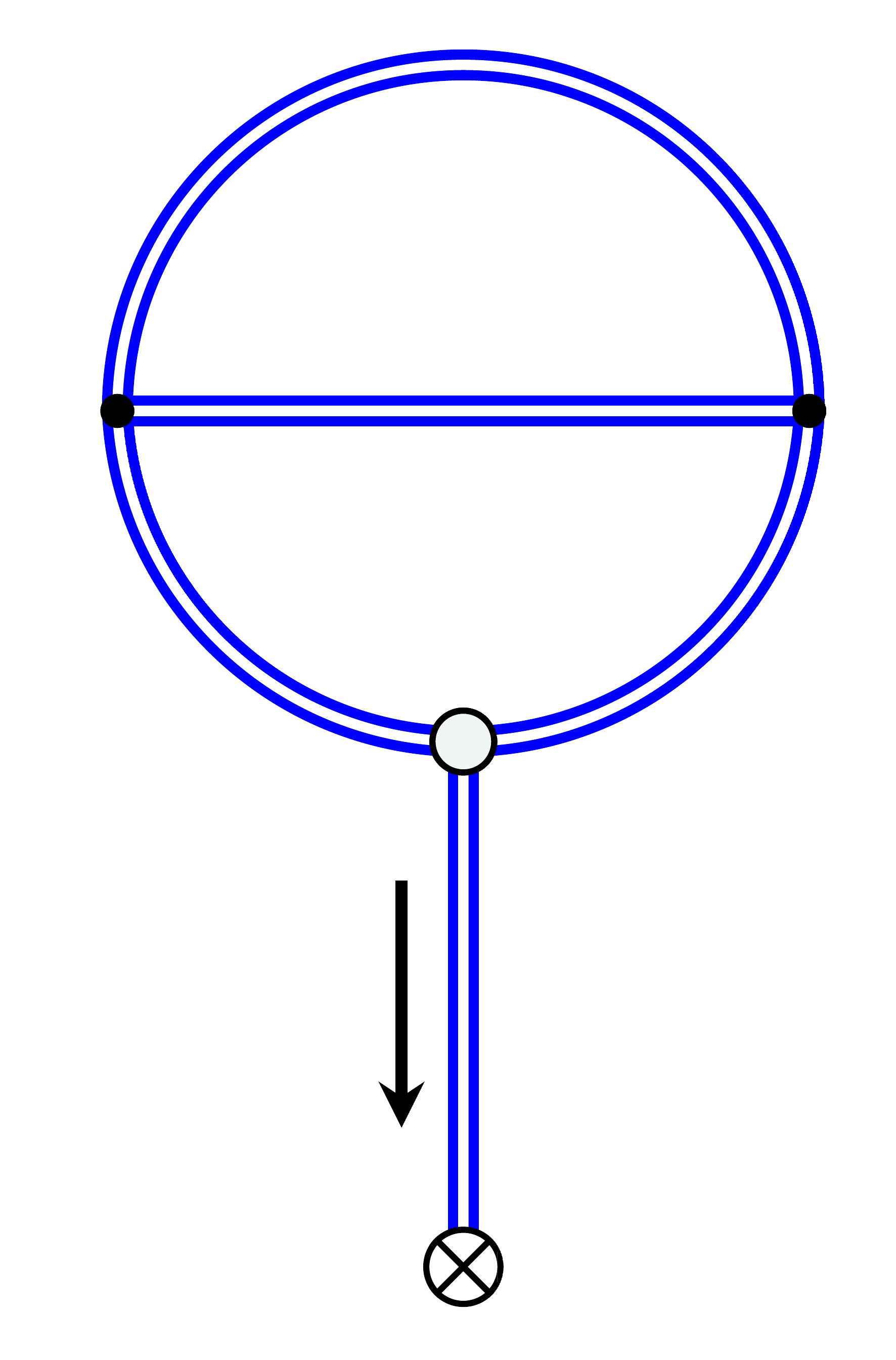}} & 
      \raisebox{0em}{%
              \includegraphics[width=.1\textwidth]%
                              {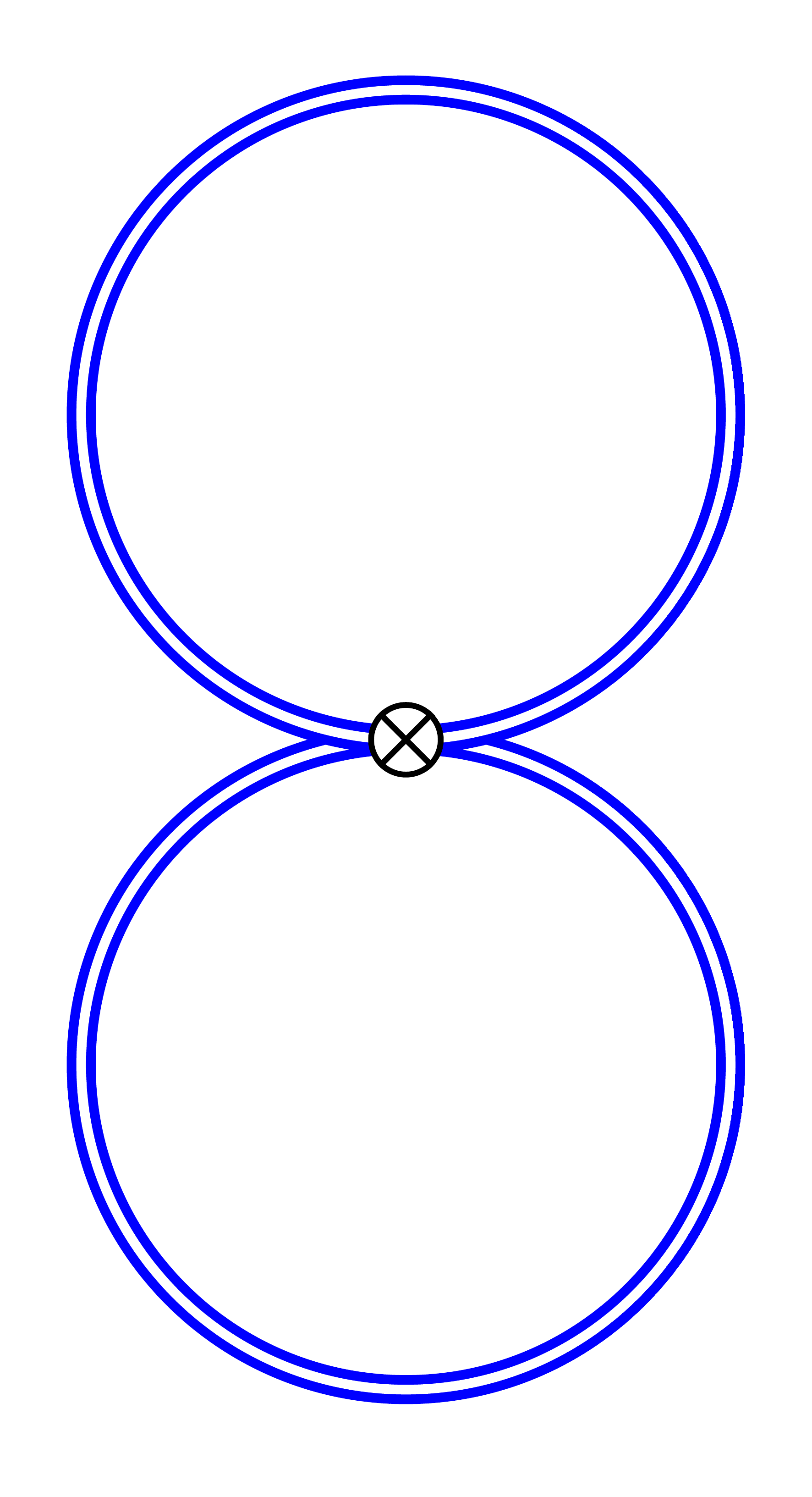}} &
      \raisebox{4em}{%
              \includegraphics[width=.1\textwidth]%
                              {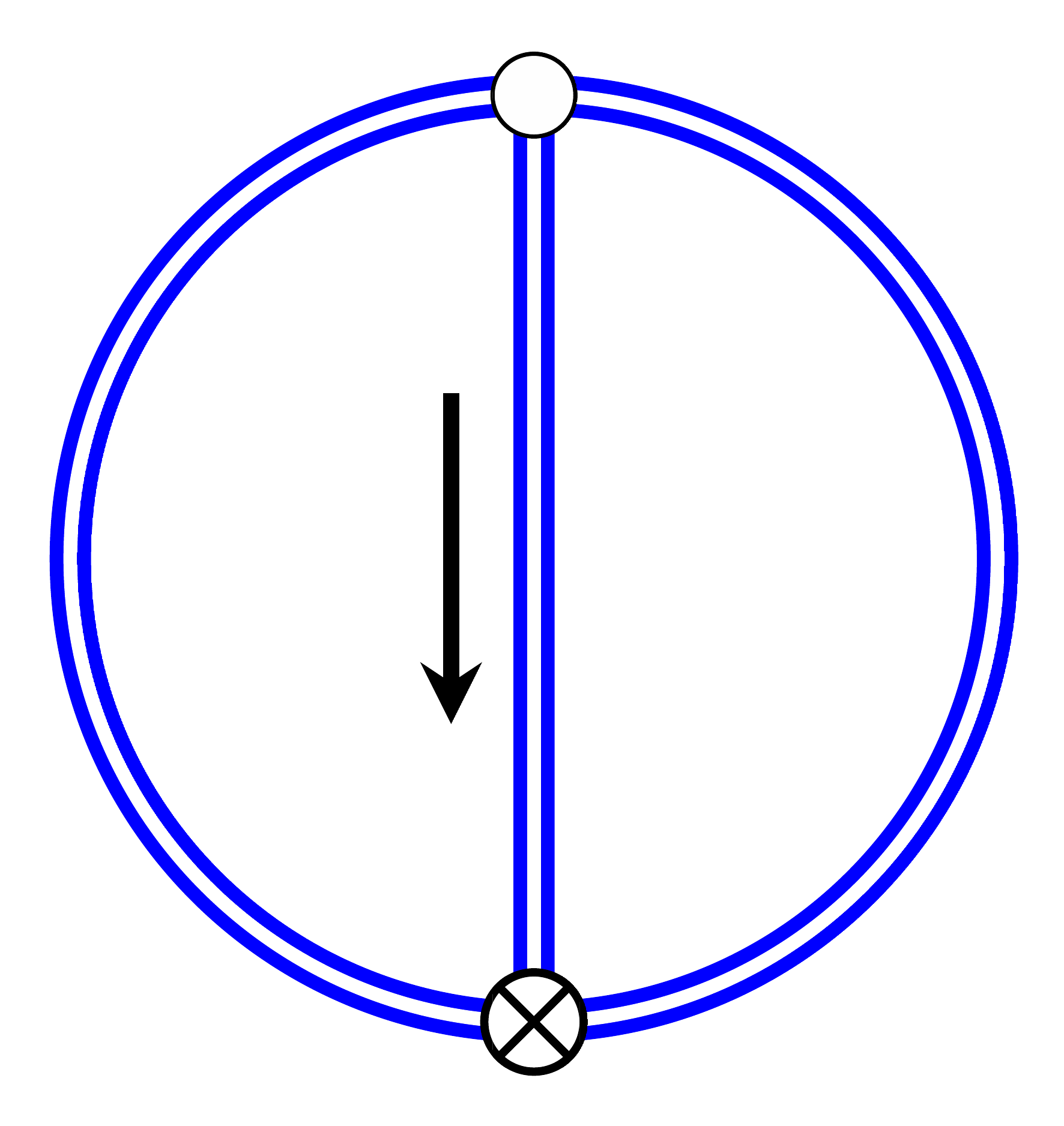}} &
      \raisebox{4.7em}{%
              \includegraphics[width=.1\textwidth]%
                              {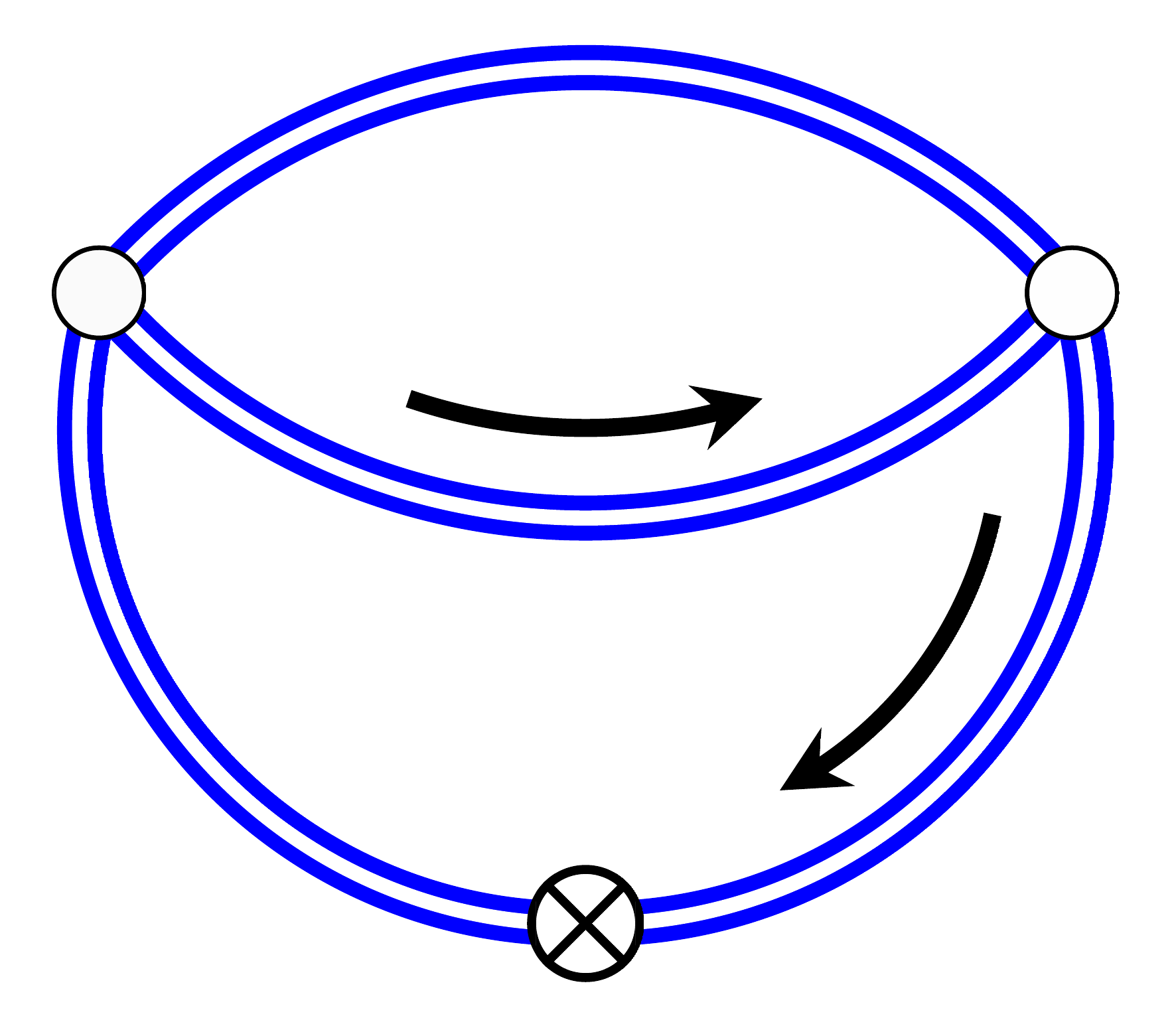}} \\
      (a) & (b) & (c) & (d)
    \end{tabular}
    \parbox{.45\textwidth}{%
      \caption[]{\label{fig:dias}\sloppy Example diagrams contributing at \nlo. (a): diagram contributing to $\mathcal{I}_1$; (b), (c), (d): present in both $\mathcal{I}_1$ and $\mathcal{I}_{\fR}$.}}
  \end{center}
\end{figure}

\prlSection{Flowed observables}We can now proceed to calculate matrix elements
of flowed operators. Specifically, we consider diffeomorphism invariant
observables which can be constructed by integrating scalars $\mathcal{\fO}_n$
over the spacetime manifold with the volume element $\sqrt{\fg}$. Taking the
\vev\ of these quantities, we define
\begin{equation}
    \mathcal{I}_n\, (t) \equiv \left\langle0\left| \intx \sqrt{\fg} \, \mathcal{\fO}_n \right|0\right\rangle \, ,
    \label{eqn:def_observables}
\end{equation}
where in the following, we consider
\begin{equation}
    \fO_1 = 1 \quad\text{and} \quad \fO_{\fR} = \fR \, .
    \label{eqn:observables}
\end{equation}
Note that these two operators are intrinsically connected by the Ricci flow equation which implies
\begin{equation}
    \partial_t \mathcal{I}_1 = - \mathcal{I}_{\fR} \, .
    \label{eq:relation}
\end{equation}
Sample \one- and \two-loop diagrams contributing to \textit{volume}
$\mathcal{I}_1$ and the \textit{integrated curvature} $\mathcal{I}_{\fR}$ are
shown in \cref{fig:dias}. After implementing the Feynman rules corresponding to
\cref{eqn:Sflow_h,eq:84:44} into the software framework described in
Ref.~\cite{Artz:2019bpr}, we calculated the resulting diagrams using the same
tool chain as applied to analogous calculations for the \qcd\ gradient
flow~\cite{Nogueira:1991ex,Harlander:1998cmq,
Nogueira:2021wfp,Gerlach:2022qnc,Klappert:2020nbg}.  All integrals can be
reduced to known one- and two-loop master integrals~\cite{Kluth:2024lar}. We
find that \cref{eq:relation} is indeed fulfilled by the bare results up to the
\two-loop level, which provides a strong check on our calculation. 

Concerning the counterterm contributions, note that the \rhs\ of \cref{eq:84:44}
is written in terms of the renormalized metric, which means that
$\delta\hat{g}_{\mu\nu}$ implicitly contains the regular counterterms $\delta
g_{\mu\nu}$. Since we calculate matrix elements of only flowed operators, the
latter are not required explicitly.  Demanding finiteness of $\mathcal{I}_1$ and
$\mathcal{I}_{\fR}$ allows us to uniquely determine the coefficients $\hat{c}_1$
and $\hat{c}_2$ of \cref{eq:292:48} in the \msbar\ scheme. We find
\begin{equation}\label{eq:274:69}
  \begin{aligned}
    \hat{c}^\text{\msbar}_1 &= -\frac{107}{30}\frac{1}{\ep}\,,\qquad
    \hat{c}^\text{\msbar}_2 = \frac{407}{30}\frac{1}{\ep}\,,
  \end{aligned}
\end{equation}
independent of the gauge parameters $\alpha$ and $\beta$,
which leads to the \msbar-renormalized results for the
\vevs:
\begin{equation}\label{eq:282:30}
  \begin{aligned}
    \mathcal{I}^{\overline{\text{MS}}}_1 &= -\frac{3\gnewton}{4\pi t} \left[1 + \frac{\gnewton}{4\pi t}  \left( 5\, \lmut + 9 \ln\frac{4}{3} + \frac{1231}{120} \right) \right] \,, \\
    t\,\mathcal{I}^{\overline{\text{MS}}}_{\hat{R}} &= -\frac{3\gnewton}{4\pi t}
    \left[1 + \frac{\gnewton}{2\pi t} \left( 5\, \lmut + 9\ln\frac{4}{3}
    + \frac{931}{120} \right) \right] \,,
  \end{aligned}
\end{equation}
where
\begin{equation}\label{eq:355:72}
  \begin{aligned}
    \lmut = \ln \mu^2 t + \EulerGamma\,,
  \end{aligned}
\end{equation}
with $\EulerGamma = 0.5772\ldots$ the Euler-Mascheroni constant. Indeed, we also
evaluated \cref{eqn:def_observables} for $\hat{\mathcal O}_n\in
\{\hat{R}_{\mu\nu\rho\sigma}\hat{R}^{\mu\nu\rho\sigma}, \hat{R}_{\mu\nu}\hat{R}^{\mu\nu}, 
\hat{R}^2, \hat{R}\Box \hat{R}\}$, and find that they are all renormalized by the 
counterterms of \cref{eq:84:44,eq:292:48,eq:274:69}. Furthermore, we observed 
that the gauge parameters $\alpha$ and $\beta$ cancel in the final 
results.

\prlSection{Ricci-flow coupling}The \vevs\ calculated above can be utilized to
define a renormalization scheme for Newton's coupling which captures power
divergences in the theory due to the dimensionful cutoff parameter $t$. Inspired
by the ``ringed scheme'' for flowed quark fields~\cite{Makino:2014taa}, we
define the \textit{\fvs} by the following all-order condition:
\begin{equation}
    \mathcal{I}_1^\text{\fvs} \equiv \mathcal{I}_1^\text{\abbrev{LO}} = -\frac{3\gnewton}{4\pi t} \,.
\end{equation}
To achieve this, we set
\begin{equation}
    \begin{aligned}
        \hat{c}_i &= \hat{c}_i^\text{\msbar} + \hat{c}_i^\text{fin}\,,\\
        \hat{c}_2^{\text{fin}} &=   10 \,\lmut + 18\log\frac{4}{3} 
        + \frac{1231}{60} - \hat{c}_1^{\text{fin}}\,,
    \end{aligned}
\end{equation}
where $c_1^\text{fin}$ is arbitrary.

\begin{figure}[!t]
    \includegraphics[width=\columnwidth]{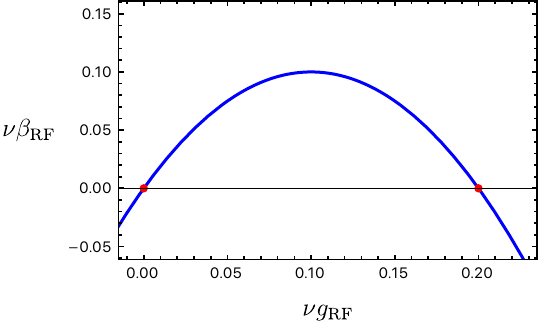}
    \caption{The $\beta$-function of the Ricci-flow  Newton coupling
    $g_\text{\RF}$ as defined in \cref{eqn:gGF_def}. 
    }
    \label{fig:betaGN}
\end{figure}

In the \fvs, the renormalized integrated curvature becomes independent of $\mu$,
\begin{equation}
    \mathcal{I}^\text{\fvs}_{\fR} = -\frac{3 \gnewton}{4\pi t^2}\left[1 - \frac{5 \gnewton}{4\pi t}\right]\, ,
    \label{eqn:VEVR}
\end{equation}
which makes it suitable for defining the gravitational coupling in the
Ricci-flow scheme:
\begin{equation}
    g_{\text{\RF}}(\mu) \equiv -\frac{t}{3\nu} 
    \mathcal{I}_{\fR}^{\text{\fvs}}\bigg|_{t=\rho/\mu^2}\,,
    \label{eqn:gGF_def}
\end{equation}
with arbitrary constants $\rho$ and $\nu$. Using the fact that $\gnewton$ is $\mu$ independent, the resulting $\beta$-function is
 \begin{equation}
    \beta_\text{\RF} \equiv \mu \frac{\partial}{\partial \mu} g_{\text{\RF}} = 2g_{\text{\RF}}\left(1 - 5\nu g_{\text{\RF}}\right)\,,
    \label{eq:beta_function_final}
\end{equation}
which is independent of $\rho$. As illustrated in
\cref{fig:betaGN}, besides the usual Gaußian
fixed point, it
exhibits a non-trivial fixed point at $g_{\text{\RF}}^* = 1/(5\nu)$. Considering
the perturbative coefficients of \cref{eqn:VEVR}, it seems reasonable to choose
$\nu\gtrsim 1/5$, which suggests that the fixed point is in the perturbative
regime. Independent of this choice, the associated critical exponent is $\theta
= 2$, rendering the coupling relevant in the ultraviolet limit $\mu \to \infty$.

\prlSection{Conclusions}We have developed a perturbative solution of the Ricci flow equation within
gravity and calculated the counterterms needed to render flowed Green's
functions finite through two-loop level. Using the \vevs\ of suitable composite
operators, we defined a Ricci-flow based renormalization scheme for Newton's
constant which exhibits a non-trival fixed point in the perturbative regime.
Further corroboration of this observation requires to study higher perturbative
orders, which should be possible with current technology, albeit with
significantly increased efforts due to the increased number of interaction
vertices and their algebraic complexity, the expected large number of
\three-loop integrals, and the increased number of counterterms that need to be
determined. On the other hand, it should be straightforward to include matter
and gauge fields in our approach, as well as a non-vanishing cosmological
constant, at least in the limit of small flow time.

Further practical applications of our approach are conceivable.  It would
certainly profit from complementary non-perturbative formulations of the Ricci
flow, for example via lattice methods in quantum gravity
\cite{Ambjorn:1991pq,Agishtein:1991cv,Ambjorn:1998xu,Ambjorn:2001cv,Loll:2019rdj}.
We also expect that the short-flow-time expansion of composite
operators~\cite{Suzuki:2013gza,Luscher:2013vga} could be helpful in
phenomenological applications such as the physics of black holes and
gravitational waves.

\begin{acknowledgments}
\prlSection{Acknowledgements}This work was supported by the Deutsche Forschungsgemeinschaft (\abbrev{DFG}) through the Walter-Benjamin program
(KL 3849/1-1, project no. 562668270), and through grant HA~2990/10-1, project no.~460791904.
\end{acknowledgments}

\bibliographystyle{apsrev4-2}
\bibliography{references}

\end{document}